\documentclass[prb,aps,preprint]{revtex4-2} 

\usepackage{amsmath}  
\usepackage{amsfonts} 
\usepackage{graphicx} 
\usepackage{enumitem}
\usepackage[normal]{subfigure}
\begin{document}


\title{Charged AdS black hole and the quantum tunneling from black hole to white hole }
\author{T.~H.~ Phat}\email{tranhuuphat41@gmail.com}
\affiliation{Vietnam Atomic Energy Commission, 59 Ly Thuong Kiet, Hanoi, Vietnam.}
\author{N.~T.~ Anh}\email{ntanh@epu.edu.vn}
\affiliation{Faculty of Energy Technology, Electric Power University, 235 Hoang Quoc Viet, Hanoi, Vietnam.}
\author{Toan T. Nguyen}\email{toannt@vnu.edu.vn}
\affiliation{Key Laboratory for Multiscale Simulation of Complex Systems, and Department of Theoretical Physics, \\
	University of Science, Vietnam National University $-$ Hanoi, 334 Nguyen Trai street, Thanh Xuan, Hanoi 100000, Vietnam.}
\author{H.~V.~Quyet \footnote{Corresponding author}}\email{hoangvanquyet@hpu2.edu.vn}
\affiliation{Department of Physics, Hanoi Pedagogical University 2, Phuc Yen,  Vinh Phuc 15000, Vietnam}


\date{\today}

\begin{abstract}
In this paper, we show that depending on the sign of the electric charge $Q$, 
the charged AdS black hole (BH) possesses two alternative facets when the cosmological constant is identified to the thermodynamic pressure $P$. 
It is discovered that: 1) the equation of state of BH corresponds to $Q > 0$ and when $Q$ changes from $Q > 0$ to $Q < 0$ we obtain the equation of state (EOS) of white hole (WH); 2) Based on the WH equation of state we found the phase transition from small to large WHs which behaves like the liquid – gas phase transition at negative temperature, $T < 0$, and, at the same time, its entropy equals with minus sign the entropy of BH. As a consequence, the latent heat $\Delta {H_{WH}}\left( T \right)$ which releases in this process equals with minus sign the latent heat of BH, too; 3) Finally, the probability of the quantum tunneling from the BH to the WH is obtained. We suggest the perspective of application of the results above to the AdS/CFT duality. 
\end{abstract}

\maketitle 

\section{Introduction}
\label{intro}
In recent years there arises great interests devoted to the quantum tunneling from black hole to white hole, for example Refs. \cite{Volovik2020, Haggard2015, de2016improved, christodoulou, bianchi2018white, barcelo2016black, haggard2015black, volovik2021black, martin2019evaporating, olmedo2017black, kedem2020black, volovik2021negative}. 
However, to our understanding, it seems that the quantum tunneling from charged AdS black hole to its partner white hole  has not been studied, so far.

In this respect, our present research will remedy this gap. To this goal, let us start from the charged AdS BH in 4-dimensional space-time whose metric reads
\begin{subequations}\label{eq1}
	\begin{eqnarray}\label{eq1a}
d{s^2} =  - f\left( r \right)d{t^2} + \frac{{d{r^2}}}{{f\left( r \right)}} + {r^2}d\Omega _{2,k}^2,
	\end{eqnarray}
	where
	\begin{eqnarray}\label{eq1b}
	f\left( r \right) = k - \frac{{2M}}{r} + \frac{{{Q^2}}}{{{r^2}}} + \frac{{{r^2}}}{{{L^2}}},
	\end{eqnarray}
\end{subequations}
and $d\Omega _{2,k}^2$ is the metric of a two-sphere ${S^2}$ of radius $1/\sqrt k $, for $ k > 0 $. The parameter $ M $ and $ Q $ related to the mass and charge of BH by corresponding factors.

The horizon is defined as the largest root of the equation
\begin{align*}
f\left( {{r_ + }} \right) = k - \frac{{2M}}{{{r_ + }}} + \frac{{{Q^2}}}{{r_ + ^2}} + \frac{{r_ + ^2}}{{{L^2}}} = 0.
\end{align*}
The temperature $ T $ and the entropy of BH are respectively defined by
\begin{align}\label{eq2}
T = \frac{1}{{4\pi }}{\left( {\frac{{df(r)}}{{dr}}} \right)_{r = {r_ + }}} = \frac{1}{{4\pi {r_ + }}}\left( {k - \frac{{{Q^2}}}{{r_ + ^2}} + \frac{{3r_ + ^2}}{{{L^2}}}} \right),
\end{align}
\begin{align}\label{eq3}
S = \pi r_ + ^2.
\end{align}
We next identify the pressure $ P $ with 
\begin{align}\label{eq4}
P = \frac{3}{{8\pi {L^2}}}.
\end{align}
In this set up $ M $ turns out to be the enthalpy of BH \cite{kubizvnak2012p}.
Introducing the topological charge $\varepsilon  = 4k$ \cite{pan2021holographic, Tian_2019} the extended first law of the BH thermodynamics \cite{articleLan, articleLan2} is found
\begin{align}\label{eq5}
dM = TdS + \omega d\varepsilon  + \phi dQ + VdP.
\end{align}
Here $\omega  = {r_ + }/8\pi$ is the conjugate potential of the topological charge, $\phi$ is the conjugate potential of charge $\phi  = Q/{r_ + }$ and the volume $ V $ conjugate to pressure $ P $ is $V = 4\pi r_ + ^3/3$. 

\section{Equations of states of black hole and white hole}
\label{sec:2}
Combining (\ref{eq1}) and (\ref{eq4}) we arrive at the equation of state (EOS) of BH
\begin{align}\label{eq6}
P = \frac{T}{{2{r_ + }}} - \frac{\varepsilon }{{32{\pi ^2}r_ + ^2}} + \frac{{{Q^2}}}{{8\pi r_ + ^4}},
\end{align}
which describes the liquid - gas phase transition in the $ P - V $ plane \cite{kubizvnak2012p} at critical point for $ Q > 0 $, $\varepsilon  > 0.$
\begin{align}\label{eq7}
{P_c} = \frac{{{\varepsilon ^2}}}{{1536{\pi ^3}{Q^2}}},\quad 
{T_c} = \frac{{{\varepsilon ^{3/2}}}}{{24\sqrt 6 {\pi ^{5/2}}Q}}, \quad 
{r_c} = \frac{{2\sqrt {6\pi } Q}}{{\sqrt \varepsilon  }},
\end{align}
which depend on the signs of two quantities $\varepsilon $ and $ Q $. Let us study the concrete cases:

a)	when $\varepsilon  > 0,Q > 0$ we have
\begin{align*}
{P_c} > 0, \quad {T_c} > 0, \quad {r_c} > 0.
\end{align*}
There is a phase transition from small to large BHs which is represented in Fig. 1a. 
Comparing with the van der Waals equation, the specific volume is chosen in \cite{kubizvnak2012p} as follows
\begin{align*}
v = 2{r_ + },
\end{align*}
which together with (\ref{eq7}) provides
\begin{align}\label{eq8}
\frac{{{P_c}{v_c}}}{{{T_c}}} = \frac{3}{8}.
\end{align}
It is very interesting that the foregoing relation coincides exactly with the one derived from the van der Waals equation 
\begin{subequations}\label{eq9}
	\begin{eqnarray}\label{eq9a}
\left( {P + \frac{a}{{{v^2}}}} \right)\left( {v - b} \right) = T,
\end{eqnarray}
where
\begin{eqnarray}\label{eq9b}
a = \frac{{3\varepsilon }}{{16{\pi ^2}}}, \quad
b = \frac{{4\sqrt {6\pi } Q}}{{3\sqrt \varepsilon  }}.
\end{eqnarray}
\begin{figure}[ht] 
	\centering 
	\subfigure[]{ 
		\includegraphics[width= 7.4cm]{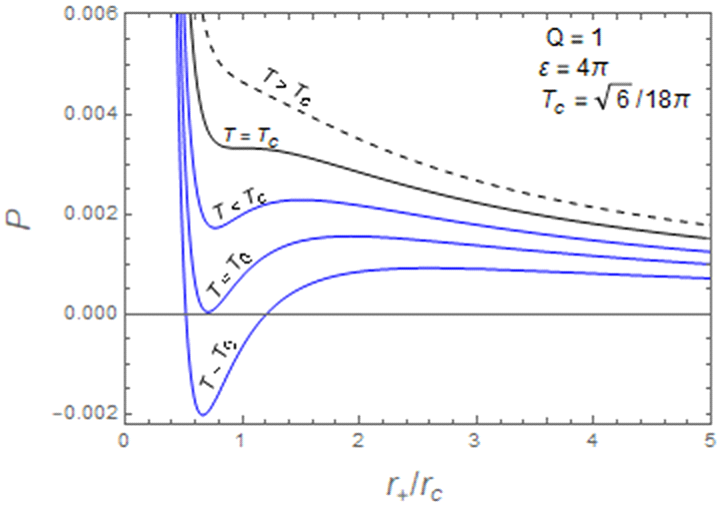}
		\label{1a}} 
	\subfigure[]{ 
		\includegraphics[width=7.4cm]{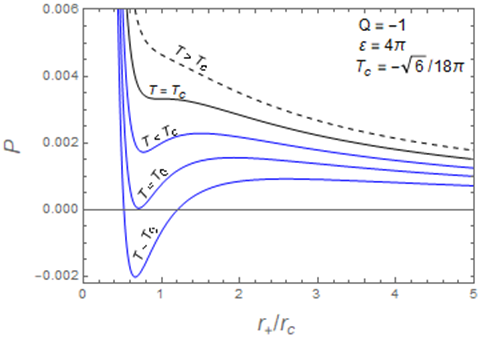}
		\label{1b}}              
\caption{The $ P - v $ diagram for $ Q > 0$ $(Q < 0) $ is plotted in Fig. 1a (Fig. 1b). 
The isotherms in Fig. 1b exhibit the phase transition from liquid to gas when the negative temperature increases from $T < {T_c}$ to $T > {T_c}$.} 
\label{1}
\end{figure}

In Fig. \ref{1a} is shown the $P - v$ diagram derived from EOS (\ref{eq6}) at different $T$ and $Q = 1$, $\varepsilon  = 4\pi $. 

b)	The case $\varepsilon  < 0$ is not acceptable because ${P_c} < 0$ and other critical quantities are imaginary. 

c)	When $\varepsilon  > 0,Q < 0$ both critical values of temperature ${T_c}$ and volume ${v_c}$ in (\ref{eq7}) are negative
\begin{align*}
{T_c} = \frac{{{\varepsilon ^{3/2}}}}{{24\sqrt 6 {\pi ^{5/2}}Q}} < 0,\quad 
{r_c} = \frac{{2\sqrt {6\pi } Q}}{{\sqrt \varepsilon  }} < 0,
\end{align*}
which can be rescued by changing
\begin{align*}
{r_c} =  - \frac{{2\sqrt {6\pi } Q}}{{\sqrt \varepsilon  }} > 0,
\end{align*}
and accepting the negative temperature
\begin{align*}
{T_c} = \frac{{{\varepsilon ^{3/2}}}}{{24\sqrt 6 {\pi ^{5/2}}Q}} < 0.
\end{align*}
At this moment the critical values 
\begin{eqnarray}\label{eq9c}
{P_c} = \frac{{{\varepsilon ^2}}}{{1536{\pi ^3}{Q^2}}},\quad 
{T_c} = \frac{{{\varepsilon ^{3/2}}}}{{24\sqrt 6 {\pi ^{5/2}}Q}},\quad 
{r_c} =  - \frac{{2\sqrt {6\pi } Q}}{{\sqrt \varepsilon  }},
\end{eqnarray}
\end{subequations}
become the critical point of the equation
\begin{align}\label{eq10}
P =  - \frac{T}{{2{r_ + }}} - \frac{\varepsilon }{{32{\pi ^2}r_ + ^2}} + \frac{{{Q^2}}}{{8\pi r_ + ^4}}.
\end{align}
We will prove later that Eq. (\ref{eq10}) is the EOS of white hole. 

Based on Eqs. (\ref{eq9c}) we obtains the interesting relation
\begin{align}\label{eq11}
\frac{{{P_c}{v_c}}}{{{T_c}}} =  - \frac{3}{8}
\end{align}
which characterizes the while hole. 

It is known that the negative temperature is well defined is condensed matter physics, for example, in \cite{1992JLTP...89..177H, 1997RvMP...69....1O} one has experimentally observed different phase transitions occurring at $ T < 0 $ in magnetic systems. 
The authors of \cite{BALDOVIN20211} assume that negative temperatures are consistent with equilibrium thermodynamics. 
Moreover, in a recent paper \cite{volovik2021negative} Volovik has shown that the negative temperature states are also possible for quantum vacuum in the relativistic quantum theories. 
The negative temperature has been discussed in the black hole physics \cite{cvetivc2018killing}. 
As a rule, the state with negative $T$ is hotter than the state with positive $ T $, since the heat flows from the negative - temperature systems to the positive- temperature systems. 
Bearing in mind the physical meaning of negative temperature we will investigate the EOS of white hole (\ref{eq10}). 

Based on Eq. (\ref{eq10}) it is easily obtained in Fig1b the isotherms in the $ P - v $ diagram at $Q = -1$, $\varepsilon  = 4\pi $. 
These isotherms express the first order phase transition from liquid to gas phases at different values of negative temperature. They look like the phase transition  in the van der Waals theory. 
However, there is a basic distinction: the liquid phase corresponds to hotter temperature, while the gas phase corresponds the cooler temperature. 
This is also the phase transition from small to large white holes which is the premise for discovering the microscopic structure inside white hole.

Writing the EOS (\ref{eq10}) in the form similar to the van der Waals equation we get
\begin{subequations}\label{eq120}
	\begin{eqnarray}\label{eq120a}
\left( {P + \frac{c}{{{v^2}}}} \right)\left( {v - d} \right) =  - kT,
	\end{eqnarray}
with $ c $ and $ d $  given by 
	\begin{eqnarray}\label{120b}
c = \frac{{3\varepsilon }}{{16{\pi ^2}}},d =  - \frac{{4\sqrt {6\pi } Q}}{{\sqrt \varepsilon  }}.
	\end{eqnarray}
\end{subequations}
It is easily seen that the critical point of Eq.(\ref{eq120}) coincides exactly with the critical point (\ref{eq9c}) of eq.(\ref{eq10}). This implies that the equality
\begin{align*}
\frac{{{P_c}{v_c}}}{{{T_c}}} =  - \frac{3}{8},
\end{align*}
is justified for these equations. 
Hence, Eqs.(\ref{eq10}) and Eq.(\ref{eq12}) are in the same corresponding class.

The Eq. (\ref{eq120}) is called the quasi – van der Waals equation for convenience. 

\section{Entropies of Black hole and white hole}
\label{sec:3}
Let us start from the EOS (\ref{eq6}) to consider the evolution of entropy ${S_{BH}}\left( T \right)$ versus $ T $ at $ Q = 1 $, $\varepsilon  = 4\pi $ and $P = {P_c} = 1/96\pi$. 
It is represented in Fig. 2a which indicates that the phase transition given in Fig. 1a is first order since the entropy is discontinuous at definite values of $ T $ corresponding to the curves $P > {P_c}$.
In accord with this we get the latent heat
\begin{align*}
\Delta {H_{BH}}\left( T \right) = T\left[ {{S_{1BH}}(T) - {S_{2BH}}\left( T \right)} \right],
\end{align*}
and its $ T $ dependence is plotted in Fig. 3a which coincides with Fig. 30 of Ref. \cite{johnston2014thermodynamic}. 
\begin{figure}[ht] 
	\centering 
	\subfigure[]{ 
		\includegraphics[width= 7.4cm]{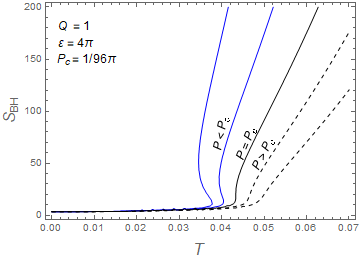}
		\label{2a}} 
	\subfigure[]{ 
		\includegraphics[width=7.4cm]{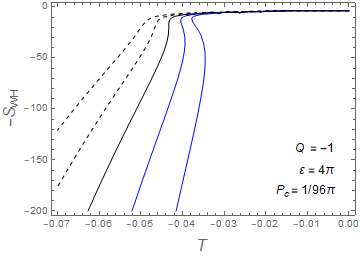}
		\label{2b}}              
	\caption{The T dependence of ${S_{BH}}\left( T \right), ( {S_{WH}}\left( T \right))$ is shown in Fig. 2a (Fig. 2b).} \label{2}
\end{figure}
\begin{figure}[ht] 
	\centering 
	\subfigure[]{ 
		\includegraphics[width= 7.4cm]{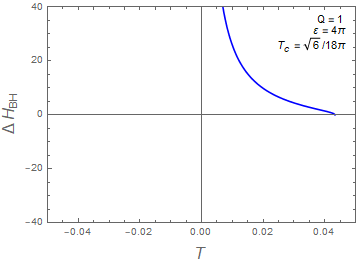}
		\label{3a}} 
	\subfigure[]{ 
		\includegraphics[width=7.4cm]{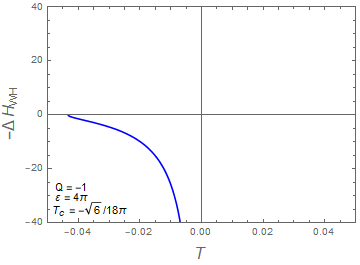}
		\label{3b}}              
	\caption{The $ T $ dependence of latent heat of black hole (white hole) is depicted in Fig. 3a (Fig. 3b).} \label{3}
\end{figure}
We next investigate the evolution of the WH entropy versus $T$ at $Q = -1$, $\varepsilon  = 4\pi ,P = {P_c} = 1/96\pi $. The obtained Fig. \ref{2b} indicates that 
\begin{align}\label{eq12}
{S_{WH}}\left( T \right) =  - {S_{BH}}\left( T \right).
\end{align}
In addition, Figs. (\ref{2a}) and (\ref{2b}) also tell that entropies of both holes obey the third law of thermodynamics: they tend to zero as temperature tend to zero 
\begin{align}
{S_{BH}}\left( 0 \right) = {S_{WH}}\left( 0 \right) = 0.
\end{align}
Finally the latent heat released in the phase transition from small to large white holes is concerned
\begin{align*}
\Delta {H_{WH}}\left( T \right) = T\left[ {{S_{1WH}}(T) - {S_{2WH}}\left( T \right)} \right].
\end{align*}
Its $ T $ dependence is shown in Fig. \ref{3a}, \ref{3b} which gives
\begin{align*}
\Delta {H_{WH}}\left( T \right) =  - \Delta {H_{BH}}\left( T \right).
\end{align*}
That is in the process of tunneling from BH to WH a quantity of energy, which is released from BH, will be absorbed by WH. 

\section{Quantum tunneling transition from black hole to white hole}   
\label{sec:4} 
The results we found in preceding sections prove that the transition from the black hole with EOS (\ref{eq6}) to the white hole with EOS (\ref{eq10}) is controlled by the electric charge $ Q $. We now apply this role of $ Q $ to calculate the probability of the quantum tunneling from the black hole to the white hole. This process consists the following steps: 

a-	The variation of Q from $ Q > 0 $ to $ Q = 0 $ leads to the transformation from the BH (\ref{eq2}) with the EOS (\ref{eq6}) to the intermediate BH (IBH)
\begin{align*}
f\left( r \right) = k - \frac{{2M}}{r} + \frac{{{r^2}}}{{{L^2}}},
\end{align*}
whose entropy is denoted by ${S_I}\left( T \right)$. 

The corresponding quantum tunneling transition can be considered as quantum fluctuation \cite{Volovik2020} and its probability is determined by the difference in the entropy between the initial and final states \cite{landau2013statistical}
\begin{align*}
P(BH \to IBH) = \exp [{S_I}\left( T \right) - {S_{WH}}\left( T \right)]
\end{align*}
b-	The variation of $ Q $ from $ Q = 0 $ to $ Q > 0 $ leads to the transformation from IBH to the white hole with EOS (\ref{eq10}). The probability of this process reads
\begin{align*}
P\left[ {IBH \to WH} \right] = \exp \left[ {{S_{WH}}\left( T \right) - {S_I}\left( T \right)} \right].
\end{align*}
Therefore, the probability of quantum tunneling from black hole to white hole equals
\begin{align*}
\begin{array}{l}
P(BH \to WH) = \exp \left[ {{S_I}\left( T \right) - {S_{BH}}\left( T \right)} \right]\exp \left[ {{S_{WH}}\left( T \right) - {S_I}\left( T \right)} \right]\\
= \exp \left[ {{S_{WH}}\left( T \right) - {S_{BH}}\left( T \right)} \right]\\
= \exp \left[ { - 2{S_{BH}}\left( T \right)} \right]
\end{array}.
\end{align*}
In summary, all the results we have obtained above prove that eq. (\ref{eq6}) is the EOS of the black hole and eq. (\ref{eq10}) is the EOS of the white hole. They are two different facets of the charged AdS black hole (\ref{eq1}), corresponding respectively to $ Q > 0 $ and $ Q < 0 $. 

\section{Conclusion}   

\label{sec:5} 
In this paper, we have studied the quantum tunneling transition from charged AdS black hole to white hole. The main results are following:

\begin{enumerate}[label=\alph*)]

\item It is shown that the electric charge $Q$ of the charged AdS black hole plays the role of a control parameter:

- When $ Q > 0 $ the system is in the black hole state, with the equation of state given by Eq. (\ref{eq6}). 

- When $ Q < 0 $ the system is in the white hole state, with Eq. (\ref{eq10}) as its equation of state.

This statement is confirmed by two equalities 
\begin{align*}
\begin{array}{l}
{S_{WH}}\left( T \right) =  - {S_{BH}}\left( T \right),\\
{T_{WH}} =  - {T_{BH}}.
\end{array}
\end{align*}

-	We have established  the quasi-van der Waals equation corresponding to the white hole equation of state.

In addition, we have obtained the relation between the latent heats of two holes 
\begin{align*}
\Delta {H_{BH}}\left( T \right) =  - \Delta {H_{WH}}\left( T \right).
\end{align*}

\item To our understanding,   the  equation of state of the white hole and its partner quasi-van der Waals equation  were never discovered before.

\item The probability of the quantum tunneling from black hole to white hole is given by
\begin{align*}
P\left( {BH \to WH} \right) = \exp \left[ { - 2{S_{BH}}} \right]
\end{align*}
A similar phenomenon also occurred in the metric of the Reissner – Nordstrom black hole: it embraces both black and white holes \cite{volovik2021black}.

\end{enumerate}

It is known that there are different approaches to consider the transition  from black hole to white hole. The general concept is to consider the white hole to be created after the decay process of black hole due to the Hawking radiation where white hole mass is smaller the black hole one \cite{barcelo2014mutiny, rovelli2019black, barcelo2017exponential}.

It is very interesting to suggest that we would have two main trends to enlarge what we have achieved:

1-	Applying the method developed in this paper to those black holes which have also the phase transition from small to large black holes such as the charged Gauss – Bonnet black holes in AdS space time \cite{unknown2013, unknown, cvetivc2002black}.

2-	Applying our method to the AdS/CFT duality may open up a new direction for searching strange matters associated with negative temperature. 
In current study, one has started from the charged AdS black hole with $ Q = 1 $ \cite{faulkner2011holographic, hartnoll2008building, hartnoll2008holographic, Horowitz_2009, horowitz2008holographic, herzog2009holographic, phat2021holographic, PhysRevD.81.041901, basu2019holographic}. 
However, we can equally choose $ Q = - 1 $ from the beginning. We then face new physical phenomena related to different phase transitions at a negative temperature such as the ones observed in experiments \cite{1992JLTP...89..177H, 1997RvMP...69....1O}.  

\bibliographystyle{apsrev4-2}


\bibliography{BHtoWH}   

\end{document}